\documentclass[final,3p,times,twocolumn]{elsarticle}
 \biboptions{comma,sort&compress}
\usepackage{here}
\usepackage{graphicx}
\usepackage{epsfig}

\usepackage{latexsym,amssymb,amsfonts,amsmath}

\def\nin{\noindent}
\def\beq{\begin{equation}}
\def\eeq{\end{equation}}
\def\bea{\begin{eqnarray}}
\def\eea{\end{eqnarray}}

\newcommand{\virg}[1]{``#1''}
\newcommand{\trpl}[1]{\vec{#1}_\perp}
\newcommand{\bitrpl}[2]{\vec{#1}_{#2\perp}}
\newcommand{\re}{\text{Re}}
\newcommand{\im}{\text{Im}}
\newcommand{\tr}{\text{Tr}}
\newcommand{\lla}{\langle\langle}
\newcommand{\rra}{\rangle\rangle}

\journal{Nucl. Phys. B (Proc. Suppl.)}

\begin{document}

\begin{frontmatter}



\title{High-energy behavior of hadronic total cross sections from lattice QCD}

\author[label1]{Enrico Meggiolaro\corref{cor1}}
\address[label1]{Dipartimento di Fisica, Universit\`a di Pisa, and
INFN, Sezione di Pisa, Largo Pontecorvo 3, I-56127 Pisa, Italy}
\cortext[cor1]{Speaker at the conference.}
\ead{enrico.meggiolaro@df.unipi.it}

\author[label2]{Matteo Giordano\corref{cor2}}
\address[label2]{Departamento de F\'\i sica Te\'orica, Universidad de Zaragoza,
Calle Pedro Cerbuna 12, E-50009 Zaragoza, Spain}
\cortext[cor2]{Supported by MICINN [CPAN project CSD2007-00042
(Consolider-Ingenio2010 program); grant FPA2009-09638].}
\ead{giordano@unizar.es}

\author[label1]{Niccol\`o Moretti}


\begin{abstract}
\noindent
By means of a nonperturbative approach to {\it soft} high-energy hadron-hadron
scattering, based on the analytic continuation of Wilson-loop
correlation functions from Euclidean to Minkowskian theory,
we shall investigate the asymptotic energy dependence of hadron-hadron
total cross sections in lattice QCD: we will show,
using best fits of the lattice data with proper functional forms
satisfying unitarity and other physical constraints, how
indications emerge in favor of a universal asymptotic high-energy
behavior of the kind $B \log^2 s$ for hadronic total cross sections.

\end{abstract}


\end{frontmatter}


\section{Introduction}
\nin
Among the oldest open problems of hadronic physics, not yet satisfactorily
solved in QCD, there is the problem of predicting hadronic total cross
sections at high energy from first principles.
Present-day experimental observations (up to a center-of-mass total energy
$\sqrt{s} = 7$ TeV, reached at the LHC $pp$ collider \cite{LHC})
seem to support the following asymptotic high-energy behavior:
$\sigma_{\rm tot}^{{}_{(hh)}} (s) \sim B \log^2 s$, with a {\it universal}
(i.e., {\it not} depending on the particular hadrons
involved) coefficient $B \simeq 0.3$ mb \cite{Blogs}. This behavior is
consistent with the well-known {\it Froissart-Lukaszuk-Martin} (FLM)
{\it theorem}~\cite{FLM}, according to which, for $s \to \infty$,
$\sigma^{{}_{(hh)}}_{\rm tot}(s) \le ({\pi}/{m_\pi^2}) \log^2 (
{{s}/{s_0}} )$, where $m_\pi$ is the pion mass and $s_0$ is an
unspecified squared mass scale. As we believe QCD to be the
fundamental theory of strong interactions, we also expect that it
correctly predicts from first principles the behavior of hadronic
total cross sections. However, in spite of all the efforts, a
satisfactory solution to this problem is still lacking.
(For some theoretical supports to the universality of $B$,
see Ref. \cite{DGN} and references therein.)

This problem is part of the more general problem of high-energy
elastic scattering at low transferred momentum, the so-called {\it soft
high-energy scattering}. As soft high-energy processes possess
two different energy scales, the total center-of-mass energy squared 
$s$ and the transferred momentum squared $t$, smaller than the typical 
energy scale of strong interactions ($|t| \lesssim 1~ {\rm GeV}^2 \ll
s$), we cannot fully rely on perturbation theory (PT). A
nonperturbative (NP) approach in the framework of QCD has been
proposed in~\cite{Nachtmann91} and further developed in~\cite{DFK}:
using a functional integral approach, high-energy hadron-hadron
elastic scattering amplitudes are shown to be governed by the
correlation function (CF) of certain Wilson loops defined in Minkowski
space~\cite{DFK}. This CF can be reconstructed by {\it analytic continuation}
from the CF of two Euclidean Wilson loops
\cite{Meggiolaro97,Meggiolaro05,GM2009}, that can be calculated
using the NP methods of Euclidean Field Theory. 
The analytic-continuation relations have allowed the NP investigation
of CFs using some analytical models, such as the \textit{Stochastic
Vacuum Model} (SVM)~\cite{LLCM2}, the \textit{Instanton Liquid Model}
(ILM)~\cite{ILM,GM2010}, the AdS/CFT \textit{correspondence}~\cite{JP},
and they have also allowed a numerical study by Monte Carlo
simulations in \textit{Lattice Gauge Theory} (LGT)~\cite{GM2008,GM2010}.

In what follows, after a brief survey of the NP approach to soft
high-energy scattering in the case of meson-meson {\it elastic}
scattering, and of the numerical approach based on LGT, we will focus
on the search for a new parameterization of the (Euclidean) CF that,
in order: $i)$ fits well the lattice data; $ii)$ satisfies unitarity
after analytic continuation; and, most importantly, $iii)$ leads to a
rising behavior of total cross sections at high energy as $B \log^2 s$,
in agreement with experimental data \cite{GMM}.
In our approach, the coefficient $B$ turns out to be \textit{universal},
i.e, the same for all hadronic scattering processes,
being related to the mass-scale $\mu$ which sets the large
impact-parameter behavior of the CF.

\section{High-energy meson-meson elastic scattering amplitude
and Wilson-loop correlation functions}

\nin
In the {\it soft} high-energy regime, the elastic scattering amplitude
${\cal M}_{(hh)}$ of two {\it mesons}, of the same mass $m$ for
simplicity, can be reconstructed from the scattering amplitude
${\cal M}_{(dd)}$ of two dipoles of fixed transverse sizes
$\vec{r}_{1,2\perp}$, and fixed longitudinal-momentum fractions
$f_{1,2}$ of the quarks in the two dipoles, after folding with squared
wave functions $\rho_{1,2}=|\psi_{1,2}|^2$ describing the interacting
hadrons~\cite{DFK},
\begin{align}
{\cal M}_{(hh)}(s,t) &= \textstyle\int
d^2\nu~\rho_1(\nu_1)\rho_2(\nu_2){\cal M}_{(dd)} (s,t;\nu_1,\nu_2)
\nonumber \\
&\equiv \lla {\cal M}_{(dd)} (s,t;1,2) \rra ,
\label{scatt-hadron}
\end{align}
where $\nu_i\!=\!(\vec{r}_{i\perp},f_i)$ denotes collectively the
dipole variables, $d^2\nu=d\nu_1d\nu_2$, $\int d\nu_i = \int
d^2\vec{r}_{i\perp}\int_0^1 df_i$, and $\int d\nu_i~\rho_i(\nu_i)=1$.
In turn, the dipole-dipole ({\it dd\/})
scattering amplitude is obtained from the (properly normalized) CF of
two Wilson loops (WL) in the fundamental representation, defined in
Minkowski spacetime, running along the paths made up of the quark and
antiquark classical straight-line trajectories, and thus forming a
hyperbolic angle $\chi \simeq \log(s/m^2)$ in the longitudinal plane. 
The paths are cut at proper times $\pm T$ as an infrared
regularization, and closed by straight-line ``links'' in the
transverse plane, in order to ensure gauge invariance; eventually,
$T\to\infty$. It has been shown in
\cite{Meggiolaro97,Meggiolaro05,GM2009} that the relevant Minkowskian
CF ${\cal G}_M(\chi;T;\vec{z}_\perp;\nu_1,\nu_2)$ ($\vec{z}_\perp$
being the {\it impact parameter}, i.e., the transverse separation
between the two dipoles) can be reconstructed, by means of
{\it analytic continuation}, from the Euclidean CF of two Euclidean WL, 
${\cal G}_E(\theta;T;\vec{z}_\perp;\nu_1,\nu_2) \!\equiv\! 
\langle {\cal W}^{{}_{(T)}}_1 {\cal W}^{{}_{(T)}}_2\rangle/(\langle
{\cal W}^{{}_{(T)}}_1 \rangle \langle {\cal W}^{{}_{(T)}}_2 \rangle )
- 1$,
where $\langle\ldots\rangle$ is the average in the sense of the
Euclidean QCD functional integral. The Euclidean WL 
${\cal W}^{{}_{(T)}}_{1,2}\!=\! N_c^{-1}\tr\{T\!\exp [-ig\oint_{{\cal C}_{1,2}}
\!{A}_{\mu}({x}) d{x}_{\mu}]\}$
are calculated on the following quark $[q]$-antiquark $[\bar{q}]$
straight-line paths, ${\cal C}_i : 
{X}_i^{{}_{q[\bar{q}]}}(\tau) = {z}_i + \frac{{p}_{i}}{m} \tau +
f^{{}_{q[\bar{q}]}}_i {r}_{i}$, 
with $\tau\in [-T,T]$, and closed by straight-line paths in the
transverse plane at $\tau=\pm T$. Here
${p}_{1,2}={m}(\pm\sin\frac{\theta}{2}, \vec{0}_{\perp},
\cos\frac{\theta}{2})$, ${r}_{i} = (0,\vec{r}_{i\perp},0)$, ${z}_i =
\delta_{i1}(0,\vec{z}_{\perp},0)$ and $f_i^{{}_{q}} \equiv 1-f_i$,
$f_i^{{}_{\bar{q}}} \equiv -f_i$. 
We define also the CFs with the infrared cutoff removed as ${\cal C}_{E,M}
\equiv\lim_{T\to\infty}{\cal G}_{E,M}$. The {\it dd}
scattering amplitude is then obtained from ${\cal C}_E(\theta;\ldots)$
[with $\theta\in(0,\pi)$] by means of analytic continuation as ($t =
-|\vec{q}_\perp|^2$) 
\begin{align}
&{\cal M}_{(dd)} (s,t;\nu_1,\nu_2) 
\!\equiv\! -i\,2s \textstyle\int d^2 \vec{z}_\perp
e^{i \vec{q}_\perp \cdot \vec{z}_\perp}
{\cal C}_M(\chi ; \vec{z}_\perp;\nu_1,\nu_2) 
\nonumber \\ \label{scatt-loop}
&= -i\,2s \textstyle\int d^2 \vec{z}_\perp
e^{i \vec{q}_\perp \cdot \vec{z}_\perp}
{\cal C}_E(\theta\to -i\chi ; \vec{z}_\perp;\nu_1,\nu_2)\, .
\end{align}
Choosing $\rho_{1,2}$ invariant under rotations and under the exchange
$f_i\to\! 1-f_i$ (see Refs.~\cite{DFK}), ${\cal C}_E$ can be
substituted in~\eqref{scatt-hadron} with the following {\it averaged}
CF:
${\cal C}_E^{ave}(\theta;|\vec{z}_{\perp}|;\hat\nu_1,\hat\nu_2) \equiv
\int\frac{d\phi_1}{2\pi} \textstyle\int\frac{d\phi_2}{2\pi}
\frac{1}{4}\{{\cal C}_E(\theta;\!\vec{z}_{\perp};\nu_1,\nu_2)+
{\cal C}_E(\theta;\!\vec{z}_{\perp};\bar\nu_1,\nu_2) +
{\cal C}_E(\theta;\!\vec{z}_{\perp};\nu_1,\bar\nu_2) + 
{\cal C}_E(\theta;\!\vec{z}_{\perp}; \bar\nu_1,\bar\nu_2)\!\}$,
where $\vec{r}_{i\perp} = |\vec{r}_{i\perp}|(\cos\phi_i,\sin\phi_i)$, 
 $\hat\nu_i=(|\vec{r}_{i\perp}|,f_i)$ and
 $\bar\nu_i=(-\vec{r}_{i\perp}, 1-f_i)$. 
Similarly, one defines the Minkowskian averaged CF,
$\mathcal{C}_M^{ave}$. As a consequence of the (Euclidean) 
{\it crossing-symmetry relations}~\cite{GM2006}, 
$\mathcal{C}_E(\pi-\theta;\vec{z}_{\perp};\nu_1,\nu_2)
\!\!=\!\!\mathcal{C}_E(\theta;\vec{z}_{\perp};\nu_1,\bar\nu_2) 
\!\!=\!\!\mathcal{C}_E(\theta;\vec{z}_{\perp};\bar\nu_1,\nu_2)$,
${\cal C}_E^{ave}$ is automatically {\it crossing-symmetric}, i.e., 
${\cal C}_E^{ave}(\pi-\theta;\ldots)={\cal C}_E^{ave}(\theta;\ldots)$.

\section{Wilson-loop correlation functions on the lattice and
comparison with known analytical results}

\nin
In Refs.~\cite{GM2008,GM2010} two of us performed a Monte Carlo
calculation of ${\cal C}_E$ in {\it quenched} QCD at lattice spacing
$a(\beta=6)\simeq 0.1\,{\rm fm}$, on a $16^4$ hypercubic lattice. We
used loops of transverse size $a$ at angles $\cot \theta\! =\! 0,\pm
1,\pm 2$. The longitudinal-momentum fractions were set to
$f_{1,2}=\frac{1}{2}$ without loss of generality~\cite{GM2010}. We
studied the configurations $\vec{z}_{\perp} \!\parallel
\!\vec{r}_{1\perp}\! \parallel\! \vec{r}_{2\perp}$ (``{\it zzz}''),
$\vec{z}_{\perp} \!\perp \! \vec{r}_{1\perp} \!\parallel\!
\vec{r}_{2\perp}$ (``{\it zyy}'') in the transverse plane, and the
averaged quantity (``{\it ave}'') defined above, for loops at
transverse distances $d \equiv |\vec{z}_\perp|/a = 0,1,2$.

Numerical simulations of LGT provide (within the errors) the true QCD
expectation for ${\cal C}_E$; approximate analytical calculations of
${\cal C}_E$ have then to be compared with the lattice data, in order
to test the goodness of the approximations involved. ${\cal C}_E$ has
been evaluated in the SVM,
${\cal C}^{\rm{}_{(SVM)}}_E\!=\! \textstyle\frac{2}{3}e^{-\frac{1}{3}
K_{\rm S}\cot\theta} + \frac{1}{3} e^{\frac{2}{3} K_{\rm S}\cot\theta} - 1$
\cite{LLCM2}, in PT,
${\cal C}_E^{\rm {}_{(PT)}}\! =\! K_{\rm p} \cot^2\theta$
\cite{BB,Meggiolaro05,LLCM2},
in the ILM,
${\cal C}^{\rm {}_{(ILM)}}_E\!=\! \frac{K_{\rm I}}{\sin\theta}$
\cite{ILM,GM2010},
and, using the AdS/CFT correspondence, for planar, strongly coupled
${\cal N}=4$ SYM at large $|\trpl{z}|$, 
${\cal C}^{\rm {}_{(AdS/CFT)}}_E = e^{\frac{K_1}{\sin\theta} +
K_2\cot\theta + K_3\cos\theta\cot\theta}-1$~\cite{JP}.
The coefficients $K_i = K_i(\trpl{z};\nu_1,\nu_2)$ are functions of
$\trpl{z}$ and of the dipole variables $\vec{r}_{i\perp}, f_i$. The
comparison of the lattice data with these analytical calculations,
performed in Ref.~\cite{GM2008} by fitting the lattice data with the
corresponding functional form, is not fully satisfactory, even though
largely improved best fits have been obtained by combining the ILM and
PT expressions into the expression ${\cal C}^{\rm {}_{(ILMp)}}_E =
\frac{K_{\rm Ip1}}{\sin\theta} + K_{\rm Ip2}\cot^2\theta $. 
Regarding the energy dependence of total cross
sections, the above analytical models are absolutely unsatisfactory,
as they do not lead to {\it Froissart-like} total cross sections at
high energy, as experimental data seem to suggest. Infact, the SVM,
PT, ILM and ILMp parameterizations lead to asymptotically constant
$\sigma_{\rm tot}^{{}_{(hh)}}$, while the AdS/CFT result leads to
power-like $\sigma_{\rm tot}^{{}_{(hh)}}$~\cite{GP2010}. 

\section{How a Froissart-like total cross section can be obtained} 

\nin
We will now introduce, and partially justify, new parameterizations of
the CF that: 
i) fit well the data;
ii) satisfy the unitarity condition after analytic continuation; and 
iii) lead to total cross sections rising as $B \log^2 s$ in the
high-energy limit \cite{GMM}. 
Regarding unitarity, from~\eqref{scatt-hadron} and \eqref{scatt-loop}
one recognizes that the quantity $A(s,|\trpl{z}|)\equiv
\lla\mathcal{C}_M(\chi;\trpl{z};1,2)\rra$ is the scattering
amplitude in impact-parameter space, which must satisfy the
{\it unitarity constraint} $|A+1|\leq 1$ (see~\cite{unitarity}). Since
$\int d\nu_i~\rho_i(\nu_i)=1$, this is the case if the following
\textit{sufficient} condition is satisfied (as we can replace
$\mathcal{C}_M\to\mathcal{C}_M^{ave}$ when averaging over the dipole
variables, a similar but weaker \textit{sufficient} condition can be
given in terms of $\mathcal{C}_M^{ave}$):
\begin{equation}\label{unitcorrdd}
|\mathcal{C}_M(\chi;\trpl{z};\nu_1,\nu_2)+1| \leq 1 \qquad
\forall \,\,\trpl{z},~\nu_1,~\nu_2.
\end{equation}
The conditions above constrain rather strongly the possible
parameterizations. For example, conditions ii) and iii) cannot be
simultaneously satisfied if the angular dependence can be factorized,
for in this case the unitarity constraint would imply
$\sigma^{{}_{(hh)}}_{\text{tot}}(\chi)\to \text{const.}$ for
$\chi\to\!\infty$. We shall \textit{assume} that the Euclidean CF can be
written as $\mathcal{C}_E=\exp K_E-1$, where
$K_E=K_E(\theta;\trpl{z};\nu_1,\nu_2)$ is a \textit{real} function
(since $\mathcal{C}_E$ is \textit{real}~\cite{GM2008}). This assumption
is rather well justified: in the large-$N_c$ expansion, $\mathcal{C}_E
\sim\mathcal{O}(1/N_c^2)$, so that $\mathcal{C}_E+1\geq0$ is certainly
satisfied for large $N_c$; all the known analytical models satisfy it; 
the lattice data of Refs.~\cite{GM2008,GM2010} confirm it. 
The Minkowskian CF is then obtained after analytic continuation:
$\mathcal{C}_M=\exp K_M-1$, with $K_M(\chi;\ldots) = K_E(\theta \to\! 
-i\chi;\ldots)$. At large $\chi$, $\mathcal{C}_M$ is expected to obey
the unitarity condition~\eqref{unitcorrdd}, which in this case
reduces to $\re K_M\leq0\,\,\, \forall \trpl{z}, \nu_1,\nu_2$.

For a {\it confining} theory like QCD, $\mathcal{C}_E$ is expected to
decay exponentially as $\mathcal{C}_E\sim\alpha\,e^{-\mu |\trpl{z}|}$
at large $|\trpl{z}|$, where $\mu$ is some mass-scale proportional to
the mass of the lightest glueball ($M_G\!\simeq\! 1.5$ GeV) or maybe
to the inverse of the so-called \textit{vacuum correlation length}
$\lambda_{vac}$ (e.g., $\mu={2}/{\lambda_{vac}}$ in the SVM), which
has been measured on the lattice~\cite{lambda-vacuum}, both in
\textit{quenched} ($\lambda_{vac}\simeq0.22$ fm) and \textit{full} QCD
($\lambda_{vac}\simeq0.30$ fm). Therefore, we should require the same
large-$|\trpl{z}|$ behavior for $K_E$, i.e., $K_E \sim e^{-\mu |\trpl{z}|}$.

Let us now assume that the leading term of the Minkowskian CF for
$\chi\!\to\!+\infty$ is of the form $\mathcal{C}_M \sim \exp\big(i\,
\beta\,f (\chi)\,e^{-\mu |\trpl{z}|}\big)-1$ [recall
$\chi\simeq\log(s/m^2)$], where $\beta\!=\!\beta(\nu_1,\nu_2)$ is a
function of the dipole variables and $f(\chi)$ is a \textit{real}
function such that $f(\chi)\to\!+\infty$ for $\chi\to\! +\infty$.
In this case, the unitarity condition~\eqref{unitcorrdd} is equivalent 
(for large $\chi$) to {$\im\beta\!\geq\!0$}. This $\trpl{z}$
dependence is expected to be valid only for large enough $|\trpl{z}|$,
but for simplicity we shall first assume that it is valid
$\forall|\trpl{z}|\!\geq\!0$. By virtue of the {\it optical theorem},
$\sigma_{\rm tot}^{{}_{(hh)}} (s)\!\sim \! s^{-1} {\rm Im} {\cal M}_{(hh)}
(s, t\!=\!0)$, we have that
$\sigma^{{}_{(hh)}}_{\text{tot}} \sim {4\pi}{\mu^{-2}} \re \lla
J(\eta,\beta) \rra$, where $J(\eta,\beta)\!\equiv\!\int_0^\infty dy \, 
y[1-\exp(i\beta e^{\eta-y})]$, with $f(\chi)=e^\eta$, and $y=\mu
|\trpl{z}|$. Expanding the exponential, integrating term by term, and
deriving with respect to $\eta$, we find 
${\partial J}/{\partial \eta}=-\textstyle\sum_{n=1}^{\infty}
{(-z)^n}/({n! n})=E_1(z) + \log(z) + \gamma$, 
for $|\arg(z)|<\pi$ ($z=-i\beta e^\eta$), where $\gamma$ is the
Euler-Mascheroni constant and $E_1(z)$ is Schl\"omilch's exponential
integral (see~\cite{GR}). Since $E_1(z) \!\sim\! e^{-z}/z$ at large
$|z|$, for $\re \, z \ge 0 \Leftrightarrow \im \beta \geq 0 $, the
asymptotic form of $\partial J/\partial\eta$ is readily obtained;
re-integrating in $\eta$ and substituting back $\eta = \log f(\chi)$,
we find $\sigma^{{}_{(hh)}}_{\text{tot}} \sim
{4\pi}{\mu^{-2}}\textstyle\lla \textstyle\frac{1}{2}\log^2
f(\chi)+\log f(\chi) (\log|\beta| + \gamma)+\dots\rra$. If one takes
$f(\chi)= \chi^p e^{n\chi}$, the resulting asymptotic behavior of
$\sigma^{{}_{(hh)}}_{\text{tot}} $ is 
\begin{equation}\label{sigmatotlead}
\vspace{-1.48pt}
\sigma^{(hh)}_{\text{tot}} \sim B \log^2 s,
\qquad \text{with:}\qquad B = \textstyle\frac{2\pi n^2}{\mu^2}.
\vspace{-1.48pt}
\end{equation}
The same result is obtained assuming the above approximation for
$\mathcal{C}_M$ only for $|\trpl{z}| \!>\!z_0 \!\gg\!
\mu^{-1},|\bitrpl{r}{i}|$: the difference in
$\sigma^{{}_{(hh)}}_{\text{tot}}$, coming from the integration of
$\mathcal{C}_M$ over the finite region $|\trpl{z}| \!<\! z_0$, 
is bounded by a constant due to the unitarity constraint.
The analysis can be repeated for $\mathcal{C}^{ave}$ without altering
any conclusion. We want to emphasize that the above result is
\textit{universal}, depending only on the mass scale $\mu$, which sets
the large-$|\trpl{z}|$ dependence of the CF, since the integration
over the dipole variables does {\it not} affect the leading term. 
The \textit{universal} coefficient $B$ is not affected by the masses of
the scattering particles: for mesons of masses $m_{1,2}$, the rapidity
becomes $\chi\sim\log(\frac{s}{m_1 m_2})$, which simply corresponds to
a change of the energy scale implicitly contained in~\eqref{sigmatotlead}.

\section{New analysis of the lattice data}

\nin
We show now three parameterizations
${\cal C}_E^{{}_{(i)}}=\exp{K_E^{{}_{(i)}}}-1$, $i=1,2,3$,
that satisfy the criteria i)--iii) listed above, together with the
corresponding estimate of the asymptotic total cross section at high
energy \cite{GMM}. We focus our analysis on the averaged CF $\mathcal{C}^{ave}$,
that is \virg{closer} to the hadronic scattering matrix ${\cal M}_{(hh)}$.
As $\mathcal{C}^{ave}$ is \textit{crossing-symmetric}, so are our
parameterizations.

\begin{figure}
\centerline{\includegraphics[width=7.5cm]{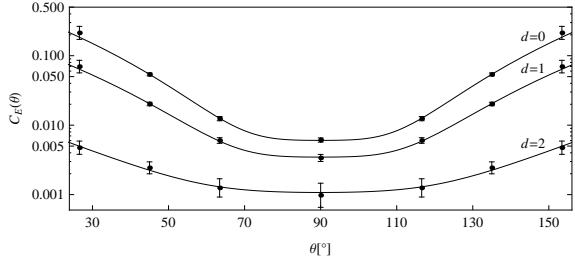}}
\caption{\scriptsize Lattice data for ${\cal C}_E^{ave}$ and best fit with
${\cal C}_E^{{}_{(3)}}$.} 
\label{corr1fit}
\end{figure}

In order to parameterize $K_E$, a possible strategy is to combine
known QCD results and variations thereof. We have then exponentiated
the two-gluon exchange and the one-instanton contribution (i.e., the
ILMp expression), adding a term which could yield a rising cross
section, e.g., a term proportional to $\cos\theta\cot\theta$, as in
the AdS/CFT result. We thus find the following parameterization: 
$K_E^{{}_{(1)}}=\frac{K_1}{\sin\theta}+K_2 \cot^2\theta + K_3
\cos\theta\cot\theta$.
Another strategy is suggested again by the AdS/CFT result: one can try
to adapt to QCD analytical expressions obtained in related models,
such as ${\cal N}\!=\!4$ SYM. Although QCD, of course, is not ${\cal N}\!=\!4$
SYM, it is sensible to assume a similar dependence on
$\theta$ (basically assuming the existence of the yet unknown gravity
dual for QCD). In this spirit, the second parameterization that we
propose is: $K_E^{{}_{(2)}} = \frac{K_1}{\sin\theta} + K_2
(\frac{\pi}{2}-\theta) \cot\theta + K_3 \cos\theta\cot\theta$.
Beside the AdS/CFT-like terms, it contains also a $\theta\cot\theta$ term.
Our last parameterization is: $K_E^{{}_{(3)}} = \frac{K_1}{\sin\theta}
+ K_2(\frac{\pi}{2}-\theta)^3 \cos\theta$. While the first term is
\virg{familiar}, the second one is not present in the known analytical
models, but it is a fact that the resulting best fit is extremely good
(see Fig.~\ref{corr1fit}).
In Table~\ref{tab:chi2} we report the values of the chi-squared
per degree of freedom ($\chi^2_{\rm d.o.f.}$) of the best fits
to the lattice data.
\begin{table}
\centering
\small
\begin{tabular}
{|l|c|c|c|}
\hline
$\chi^2_{\rm d.o.f.}$ & $d=0$ & $d=1$ & $d=2$\\
\hline
Corr 1 & 2.81 & 1.25 & 0.05 \\
Corr 2 & 0.55 & 0.31 & 0.05 \\
Corr 3 & 0.17 & 0.11 & 0.10 \\
\hline
\end{tabular}
\caption{\scriptsize Chi-squared per degree of freedom for a best fit with the
indicated function.}
\label{tab:chi2}
\end{table}

In the three cases, the unitarity condition $\re K_M^{{}_{(i)}}\le 0$
is satisfied if $K_2\geq0$: this is actually the case for our best
fits (within the errors). The leading term after analytic continuation
is of the form $\chi^pe^{\chi}$ which, according
to~\eqref{sigmatotlead}, leads to $\sigma^{{}_{(hh)}}_{\text{tot}}
\sim B \log^2 s$. The value of $B=2\pi/\mu^2$, obtained through a fit
of the coefficient of the leading term with an exponential function
$\sim e^{-\mu |\trpl{z}|} $ over the available distances, is found to
be compatible with the experimental result (within the large errors)
in all the three cases (see Table~\ref{tab:lambdavac}). However, this
must be taken only as an estimate, as lattice data are available only
for small $|\trpl{z}|$. 

\begin{table}
\centering
\small
\begin{tabular}{|l|c|c|c|}
\hline
& $\mu$ (GeV) & $\lambda=\frac{1}{\mu}$ (fm) & $B=\frac{2\pi}{\mu^2}$ (mb) \\
\hline
Corr 1 & $4.64(2.38)$ & $0.042^{+0.045}_{-0.014}$ & $0.113^{+0.364}_{-0.037}$ \\
Corr 2 & $3.79(1.46)$ & $0.052^{+0.032}_{-0.014}$ & $0.170^{+0.277}_{-0.081}$ \\
Corr 3 & $3.18(98)$ & $0.062^{+0.028}_{-0.015}$ & $0.245^{+0.263}_{-0.100}$ \\
\hline
\end{tabular}
\caption{\scriptsize Mass-scale $\mu$, \virg{decay length} $\lambda=1/\mu$ and
the coefficient $B=2\pi/\mu^2$ obtained with our parameterizations.}
\label{tab:lambdavac}
\end{table}

\section{Conclusions}
\nin
We have shown how a {\it universal} and {\it Froissart-like} hadron-hadron
total cross section at high energy can emerge in QCD, and we have found
indications for this behavior from the lattice. The functional integral
approach provides the \virg{natural} setting for achieving this result,
since it encodes the energy dependence of hadronic scattering amplitudes
in a single \textit{elementary} object, i.e., the loop-loop CF.

\end{document}